\documentclass[12pt]{JHEP3} 
\usepackage{amsmath}
\usepackage{epsfig}
\input epsf.tex
\usepackage{amssymb,amsfonts}

\newcommand{\be}{\begin{equation}}
\newcommand{\ee}{{\end{equation}}}
\newcommand{\ba}{\begin{eqnarray}}
\newcommand{\ea}{{\end{eqnarray}}}

\newcommand{\E}{{\cal E}}

\renewcommand{\H}{{\cal H}}

\newcommand{\ie}{{\it i.e.~}}

\newcommand{\Zop}{\mathbb{Z}}

\newcommand{\g}{\mathfrak{g}}
\newcommand{\h}{\mathfrak{h}}
\newcommand{\rk}{{\rm rk}}

\newcommand{\nn}{{\nonumber}}

\def\sqr#1#2{{
\vcenter{\vbox{\hrule height.#2pt
\hbox{\vrule width.#2pt height#1pt \kern#1pt
\vrule width.#2pt}
\hrule height.#2pt}}}}
\boldmath
\title{The D-branes of SU(n)}
\unboldmath
\author{Matthias R.\ Gaberdiel$^{1}$, Terry Gannon$^{2}$ and 
Daniel Roggenkamp$^{1}$
\\
$\ $ \\
$\ $ \\
$^1$Theoretische Physik, ETH-H\"onggerberg\\
\ CH-8093 Z\"urich, Switzerland\\
\ Email: \email{gaberdiel@itp.phys.ethz.ch}, 
\email{roggenka@phys.ethz.ch} \\
$\ $ \\
$^2$Department of Mathematical Sciences, University of
Alberta\\
\ Edmonton, Alberta, Canada, T6G 2G1\\
\ Email: \email{tgannon@math.ualberta.ca}
}

\abstract{D-branes that appear to generate all the K-theory
charges of string theory on {\rm SU(n)} are constructed, and their
charges are determined.} 

\preprint{hep-th/0403271}


\begin{document}

\section{Introduction}

It is widely believed that the charges of D-branes can be described
in terms of (twisted) K-theory \cite{mm,WittenK,moore}. For example,
for the case of string theory on the simply connected group $G$, the
charge group is conjectured to be the twisted K-group of $G$
(for more details see for example \cite{freed1,braun}). 
Modulo some technical assumption, this twisted K-group has been
calculated in \cite{braun} (see also \cite{freed2,fht}) to be  
\begin{equation}\label{chargegroup}
^{k+h^{\vee}}K^*(G)\cong \Zop_{d(G,k)}^{\oplus m}\,,
\quad m=2^{\rk(\bar\g)-1}\,,
\end{equation}
where $h^\vee$ is the dual Coxeter number of $G$, $k$ is the level of
the underlying WZW model, and $d(G,k)$ is the integer 
\begin{equation}
d(G,k) = \frac{k+h^\vee}{{\rm gcd}(k+h^\vee,L)}\,,
\end{equation}
where $L$ only depends on $\bar{\g}$, the finite dimensional Lie
algebra associated to $\g_k$. The summands $\Zop_{d(G,k)}$ are 
equally divided between even and odd degree if $\rk(\bar\g)>1$. For 
$\bar\g=\mathfrak{su}(2)$ the only summand is in even degree.
\smallskip

On the other hand, D-branes can be constructed in terms of the
underlying conformal field theory, and it should be possible to
determine their charges using this microscopic description. In
particular, it was shown in \cite{fs} how the charges for branes
that preserve the affine algebra $\g_k$ (up to an automorphism) can in
principle be calculated. For the D-branes that preserve the full
algebra without any automorphism, the charges were then determined for
SU(N) \cite{fs,mms}, and later for all  simply-connected Lie groups
\cite{bouwknegt}, and it was found that they account precisely for one
summand $\Zop_{d(G,k)}$. More recently, the charges of the D-branes that
preserve the affine algebra up to an outer automorphism were
determined \cite{gg}; whenever these twisted D-branes exist, they also
contribute one summand $\Zop_{d(G,k)}$. These D-branes therefore only
account at most for two of the summands $\Zop_{d(G,k)}$ in 
(\ref{chargegroup}); a CFT description of D-branes carrying charges
which lie in the remaining summands is still missing.

In \cite{mms} it was suggested that at least some of the remaining
D-branes could be described by branes that only preserve the affine
algebra $\h$ associated to a Cartan subalgebra $\bar\h\subset\bar\g$,
together with the symmetry of the coset $\g_k / \h$. 
The affine algebra  
$\h$ is obviously just the algebra of $r=\rk(\bar\g)$ free bosons, and
one can choose $r$ Neumann or Dirichlet boundary conditions
independently from one another. There are therefore something like
$2^r$ different gluing conditions that can be generated in this 
manner, and one may expect that D-branes with these gluing conditions  
can account for the full charge
group (\ref{chargegroup}). The details of this proposal were however
not worked out. Furthermore, it was not clear how to calculate
the charges that are produced by these branes.

In this paper we shall give a detailed construction of these branes,
extending techniques given in \cite{qs,qs1}.
Furthermore we shall explain how to determine the associated
charges, and we shall find that they account precisely for the missing
summands in (\ref{chargegroup}). We shall only consider the case of 
$A_n$ in this paper, but we expect that our construction
will generalise uniformly to the other cases as well.

Most of our discussion will be phrased in terms of the purely bosonic
theory, but it is easy to see that the branes we construct
actually preserve the $N=1$ worldsheet supersymmetry (provided we
choose the corresponding boundary conditions for the fermions); they 
therefore give rise to spacetime supersymmetric branes. 

The paper is organised as follows. In the next section we 
review the WZW models, and in particular explain how the space of
states can be written in terms of representations of the coset algebra
$\g_k / \h$ and the algebra $\h$. The boundary states are constructed
in section~3, and in section~4 we argue that these branes actually
generate the full K-theory group.

\section{Some details about WZW-models}
\setcounter{equation}{0}

String theory on a group manifold $G$ is described in terms of
representations of the affine Lie algebra $\g$ at level $k$. For the
situation where the group manifold is simply-connected (a systematic
analysis of the D-brane charges on non simply-connected manifolds was
recently started in \cite{ggone}), the full spectrum of the theory is
then
\begin{equation}\label{spectrum}
\H = \bigoplus_{\lambda\in P^k_+} \H_\lambda \otimes
\bar\H_{\lambda^\ast} \,,
\end{equation}
where the sum runs over all integrable highest weight representations
$\lambda$ of $\g_k$, and the representation for the right-movers is
conjugate to the representation of the left-movers. (This theory is
therefore sometimes referred to as the `charge-conjugation' theory.)

We are interested in constructing boundary conditions
that only preserve the 
subalgebra $\h$ that is associated to a Cartan subalgebra
$\bar\h\subset \bar\g$, as well as the corresponding coset algebra
$\g_k / \h$. For this purpose it is convenient to decompose the space
of states (\ref{spectrum}) in terms of representations of these
algebras. In order to do so, we need to introduce some notation. 

By $\bar\g$ we mean the finite dimensional simple Lie algebra
corresponding to $\g_k$. We assume for convenience that $\bar\g$ is
simply laced; the generalisation to the non simply laced case is
straightforward. The root lattice of $\bar\g$ is denoted by
$\Lambda_R$, and the weight lattice is 
$\Lambda_W\cong \Lambda_R^\ast$. A basis for the root lattice is given
by the simple roots $\alpha_i$, $i=1,\ldots,\rk(\bar\g)$, all of whose
length squares are equal to $2$. The corresponding dual basis of the
weight lattice is given by the fundamental weights $\Lambda_i$, \ie\  
$\Lambda_i \cdot \alpha_j = \delta_{ij}$. For the case of 
$\bar\g=\mathfrak{su}(n+1)$ the inner products of the 
fundamental weights are 
\begin{equation}
\Lambda_i \cdot \Lambda_j = \frac{i (n+1-j)}{n+1} \,,
\end{equation}
where $i\leq j$. (For a general introduction to these matters see for
example \cite{fsbook}.)

The free bosons that make up $\h$ give rise to an extended symmetry
algebra\footnote{The symmetry algebra is extended
by vertex operators associated to elements in $k\Lambda_R$.
The latter have integer conformal weight since the conformal weight of
the state associated to $\mu\in\Lambda_W$ in the free boson theory is 
$\mu^2/2k$.}, whose representations are parametrised by 
$P_+^\h = \Lambda_W / k \Lambda_R$; the corresponding
representation spaces decompose as 
\begin{equation}
\widehat{\H}^\h_{\mu} = \bigoplus_{\delta\in k\Lambda_R} 
\H^{\h}_{\mu+\delta} \,, 
\end{equation}
where $\H^{\h}_{\mu+\delta}$ is the Fock space that is generated by
the action of the $\rk(\bar\g)$ bosonic oscillators from a ground state
$|\mu+\delta\rangle$ that is an eigenvector of the oscillator zero
modes. 

For the case of $\bar\g=\mathfrak{su}(n+1)$ the modular
$S$-matrix of this extended symmetry algebra is then
\begin{equation}
S^\h_{\mu \mu'} = \frac{1}{\sqrt{n+1} \, k^{n/2}} \,
e^{2\pi {\rm i} \frac{\mu \cdot \mu'}{k}} \,,
\end{equation}
and it leads to the usual fusion rules
\begin{equation}
{N^\h_{\mu_1 \mu_2}}^{\mu_3} 
= \delta^{(k\Lambda_R)}_{\mu_3,\mu_1+\mu_2} \,.
\end{equation}

The representations of the coset algebra $\g_k / \h$ are labelled by
pairs $(\lambda,\mu)$, where $\lambda\in P^k_+$, $\mu\in P_+^\h$, and   
$\lambda-\mu\in\Lambda_R$. Not all of these pairs of representations
are inequivalent. For example, for $\g_k=\widehat{\mathfrak{su}}(n+1)_k$
two pairs $(\lambda_1,\mu_1)$ and $(\lambda_2,\mu_2)$ define the same
coset representation if and only if 
\begin{equation}\label{simplecurrents}
\lambda_1 = J^l \lambda_2 \qquad {\rm and}\qquad
\mu_1 = \mu_2 + k l J' \qquad
({\rm mod}\, k \Lambda_R) \,,
\end{equation}
where $J$ is the simple current acting on affine weights 
$(\lambda_0;\lambda_1,\ldots,\lambda_{n})$
of ${\widehat{\mathfrak{su}}}(n+1)_k$
by $J(\lambda_0;\lambda_1,\ldots,\lambda_{n})=
(\lambda_{n};\lambda_0,\ldots,\lambda_{n-1})$; the corresponding simple
current of the free bosonic theory acts on the weights by addition
of the $n$th fundamental weight $J'=\Lambda_n$.
We denote the group of these field identifications by $G_{id}$; by
construction it has order $n+1$.   

Let us denote by $[\lambda,\mu]$ the equivalence class of such pairs, 
\ie\ the orbit $[\lambda,\mu] = G_{id} (\lambda,\mu)$. Then the
$S$-matrix of the coset theory is simply 
\begin{equation}
S^{\g/\h}_{[\lambda,\mu] \, [\lambda',\mu']} = (n+1) \,
S^{\g}_{\lambda \lambda'} \, \bar{S}^\h_{\mu \mu'} \,.
\end{equation}
It is easy to check that this matrix is well defined on the
equivalence classes, and that it is unitary; in particular,
this implies that the above field identifications are in fact
the only field identifications. 

For each $\lambda\in P^k_+$ we can decompose the corresponding
space of states as 
\begin{equation}\label{decompc}
\H_\lambda = \bigoplus_{\mu\in\Lambda_W} 
\H_{[\lambda,\mu+k\Lambda_R]} \otimes \H^\h_{\mu}\,,
\end{equation}
where the sum runs over all $\mu$ for which
$\lambda-\mu\in\Lambda_R$, and 
$\H_{[\lambda,\mu+k\Lambda_R]}$ is the
representation of the coset algebra. The total space of states
therefore has the decomposition 
\begin{equation}\label{decomp}
\H  = \bigoplus_{\lambda\in P^k_+}
\bigoplus_{\mu,\bar\mu\in\Lambda_W} \, 
\left( \H_{[\lambda,\mu+k\Lambda_R]} \otimes
\bar\H_{[\lambda^\ast,\bar\mu+k\Lambda_R]} \right)
\otimes
\left(\H^\h_{\mu}\otimes \bar\H^{\h}_{\bar\mu} \right) \,,
\end{equation}
where again $\lambda-\mu,\lambda^\ast-\bar\mu\in\Lambda_R$.

\section{Boundary conditions}
\setcounter{equation}{0}

As we have mentioned before, we are interested in constructing
boundary states of the above WZW model that only preserve in general
the subalgebra $\h$, as well as the coset algebra $\g_k/\h$.
Obviously the usual `untwisted' and `twisted' boundary states (whose
construction is well understood) are special examples of such
boundary states. We want to generalise their construction by 
changing the gluing conditions for the free bosons that make up
the Cartan subalgebra $\bar\h$; in particular, we want to consider
different combinations of Neumann and Dirichlet boundary conditions
for these bosons. 

More specifically, the usual D-branes of the WZW model are
characterised by the gluing conditions
\begin{equation}\label{currentglue}
\left( J^a_n + \omega (\bar{J}^a_{-n}) \right) \, 
|\!| \omega \rangle\!\rangle = 0 \,,
\end{equation}
where $\omega={\rm id}$ for the untwisted branes, and $\omega$ is the 
non-trivial outer automorphism for the case of the twisted
branes. (For $A_n$ the outer automorphism is simply charge
conjugation, $\omega=C$.)

In terms of the subalgebra $\g_k/\h$, these gluing
conditions are 
\begin{equation}\label{cosetglue}
\left( S_n - (-1)^{h_S} \omega(\bar{S}_{-n}) \right) \, 
|\!| \omega \rangle\!\rangle = 0 \,,
\end{equation}
where $S_n$ denote the modes of a field in the coset algebra 
$\g_k/\h$, and $\omega$ is the induced automorphism on this algebra. 
In particular, there are therefore (at least) two different gluing
conditions that can be imposed on the coset algebra. 

The gluing conditions for the free bosons that make up $\h$ are
\begin{equation}\label{bosonglue} 
\left( H^i_n + \sigma^{ij} \bar{H}^j_{-n} \right) \,  
|\!| \sigma, \omega \rangle\!\rangle = 0 \,,
\end{equation}
where $\sigma$ is an orthogonal matrix; for $\omega={\rm id}$, 
$\sigma={\bf 1}$, while for $\omega=C$, $\sigma=-{\bf 1}$. We want to 
generalise this construction by considering more general $\sigma$ for
each given choice of $\omega$, \ie\ we want to construct the boundary
states that are characterised by (\ref{cosetglue}) and
(\ref{bosonglue}), but that do not necessarily satisfy
(\ref{currentglue}). 

For the following it will be convenient to choose a suitable
orthogonal basis for the Cartan subalgebra $\bar\h$, and to restrict
the construction to those $\sigma$ that are diagonal with respect to
this basis. As before, let $\Lambda_i$, $i=1,\ldots, n$ be the
fundamental weights of ${\mathfrak{su}}(n+1)$. We define
\begin{eqnarray}
\tilde\Lambda_n & = & \Lambda_n \\
\tilde\Lambda_j & = &  \Lambda_j - \frac{j}{j+1} \Lambda_{j+1} \,,\qquad
 1\leq j<n\,.
\end{eqnarray}
One then easily checks that these weights are pairwise orthogonal,
\begin{equation}
\tilde\Lambda_i \cdot \tilde\Lambda_j = \delta_{ij} \, 
\frac{j}{j+1}
\,. 
\end{equation}
If one considers the filtration of algebras
$\mathfrak{su}(2)\subset \mathfrak{su}(3) \cdots \subset 
\mathfrak{su}(n)\subset \mathfrak{su}(n+1)$, where 
$\mathfrak{su}(l+1)$ is generated by the first $l$ fundamental roots,
then $\tilde\Lambda_j$ is just the $j$th fundamental weight of
$\mathfrak{su}(j+1)$. This is a consequence of the fact that
\begin{equation}\label{allam}
\alpha_j \cdot \tilde\Lambda_l = \left\{
\begin{array}{cl}
0 \quad & \hbox{if $j<l$ or $j>l+1$} \\
1 \quad & \hbox{if $j=l$} \\
-\frac{l}{l+1} \quad & \hbox{if $j=l+1$.}
\end{array}
\right.
\end{equation}
We shall consider $\sigma$ that are diagonal in this basis. Since
$\sigma$ is orthogonal, the possible eigenvalues of $\sigma$ are just
$\pm 1$. We choose the convention that 
$\sigma\tilde\Lambda_j = s_j \tilde\Lambda_j$ 
where $s_j=\pm 1$. The action of $\sigma$ on a general
$\bar\mu\in\Lambda_W$ is then
\begin{equation}\label{Rmuex}
\sigma\, \bar\mu = \sum_j s_j \frac{j+1}{j} 
(\tilde\Lambda_j \cdot \bar\mu) 
\, \tilde\Lambda_j \,,
\end{equation}
where we have used that $\bar\mu$ can be written as 
\begin{equation}\label{muex}
\bar\mu = \sum_j \frac{j+1}{j} (\tilde\Lambda_j \cdot \bar\mu)
\, \tilde\Lambda_j \,.
\end{equation}

The first step of the construction consists of identifying the
Ishibashi states that satisfy (\ref{cosetglue}) and 
(\ref{bosonglue}). The analysis will depend on whether $\omega$ is the
trivial automorphism or charge conjugation, and we will therefore have
to consider these two cases in turn.

\subsection{The untwisted construction}

Let us first consider the case where $\omega={\rm id}$. It then
follows from (\ref{decomp}) that we get an Ishibashi state for every
$(\lambda,\mu,\bar\mu)$ for which
$\mu-\lambda,\bar\mu-\lambda^\ast\in\Lambda_R$ and
\begin{equation}
\mu = - \sigma \bar\mu \,, \qquad \qquad
\mu = - \bar\mu \quad (\hbox{mod}\, k \Lambda_R) \,.
\end{equation}
By combining these two conditions we therefore get precisely one
Ishibashi state for each $\bar\mu\in\Lambda_W$ for which 
\begin{equation}\label{condun}
({\bf 1} - \sigma)  \, \bar\mu \in k \Lambda_R \,.
\end{equation}
If $\bar\mu$ satisfies (\ref{condun}) then 
$\mu:= -\sigma\bar\mu\in\Lambda_W$. Furthermore, since 
$\mu+\bar\mu\in \Lambda_R$, $\lambda-\mu\in\Lambda_R$. 
Let us denote the corresponding Ishibashi state by
$|\lambda,\bar\mu\rangle\!\rangle$. Our ansatz for the boundary state
is then   
\begin{equation}\label{ansatz}
|\!| \sigma,\nu \rangle \!\rangle = \sqrt{|\Gamma_\sigma|}\,
\sum_{\lambda\in P_+^k} 
\frac{S_{\lambda \nu}}{\sqrt{S_{\lambda 0}}} \,
\sum_{\bar\mu+\lambda\in\Lambda_R}{}^{\!\!\!\!\prime}
\, e^{{\rm i}\, \theta \cdot \bar\mu}
\, 
|\lambda,\bar\mu\rangle\!\rangle \,,
\end{equation}
where the last sum is restricted to the solutions of (\ref{condun})
and $|\Gamma_\sigma|$ is the order of a finite abelian group
$\Gamma_\sigma$ that will be defined below.  
The parameter $\theta$ describes the position or Wilson line, and we
shall set $\theta=0$ from now on. The $S$-matrix here is the
$S$-matrix of $\g_k$, and we have used that
$\bar\mu-\lambda^\ast\in\Lambda_R$ if and only if  
$\bar\mu+\lambda\in\Lambda_R$. 

In order to calculate the overlap between these boundary states, it is
important to understand how to characterise the solutions of
(\ref{condun}). Using the explicit formulae (\ref{Rmuex}) and 
(\ref{muex}) we can rewrite (\ref{condun}) as 
\begin{equation}\label{condun1}
\frac{1}{k} \sum_j (1-s_j) \frac{j+1}{j} \,
(\tilde\Lambda_j\cdot \bar\mu) \, \tilde\Lambda_j \in \Lambda_R \,.
\end{equation}
A vector $\bar\gamma$ is in the root lattice if both
$\Lambda_n\cdot\bar\gamma\in\Zop$, and if $\bar\gamma$ is in 
the weight lattice, \ie\ if $\alpha_j\cdot \bar\gamma\in\Zop$ for all
$j$. Dotting the above equation with $\Lambda_n$ we therefore obtain
the constraint that 
\begin{equation}\label{ncond}
(1-s_n) \, \Lambda_n \cdot \bar\mu \in k\, \Zop \,.
\end{equation}
If we define the $n$-ality of a weight $\mu\in\Lambda_W$ by 
$t(\mu) = \sum_{j} j \mu_j$, then (\ref{ncond}) 
can simply be written as 
\begin{equation}
(1-s_n) \, t(\bar\mu) \in k\, (n+1)\, \Zop \,.
\end{equation}
If $s_n=+1$, this condition is empty, but if $s_n=-1$ it implies that
there are $\lambda\in P^k_+$ for which no Ishibashi state 
$|\lambda,\bar\mu\rangle\!\rangle$ exists. This suggests that the 
construction will break down for $s_n=-1$, and this is indeed what
will become apparent below. (If $s_n=-1$ one has to use the twisted
construction instead that will be described in the following
subsection.) For now we therefore assume that $s_n=+1$. 

This leaves us with analysing the condition that (\ref{condun1}) is
actually a weight. Dotting the equation by $\alpha_j$ and using 
(\ref{allam}) we obtain
\begin{equation}\label{3.16}
\frac{1}{k} \left[ 
(1-s_j) \frac{j+1}{j} \, (\tilde\Lambda_j \cdot\bar\mu) - 
(1-s_{j-1}) \, (\tilde\Lambda_{j-1} \cdot\bar\mu) \right] \in \Zop
\,. 
\end{equation}
Since $\alpha_j=2\Lambda_j - \Lambda_{j-1}-\Lambda_{j+1}$ we can
rewrite this condition as 
\begin{equation}\label{3.17}
\left[ (1-s_j)\, \alpha_j + (s_{j-1}-s_j) \, \tilde\Lambda_{j-1}
\right] \cdot \bar\mu \in k\, \Zop \,.
\end{equation}
This condition has to hold for all $j=1,\ldots, n$. As is explained in
the appendix, one can construct a projector $P_\sigma$ that projects
any state in $\H$ onto the components for which $\bar\mu$ satisfies 
(\ref{condun}). This projector can be written as 
\begin{equation}\label{condun2}
P_\sigma = \frac{1}{|\Gamma_\sigma|}\, 
\sum_{v\in\Gamma_\sigma} 
\exp\left[2\pi {\rm i}\, v \cdot \bar{H}_0 \right] \,,
\end{equation}
where $\Gamma_\sigma$ is a finite abelian group that is a quotient of
a lattice of shift vectors by roots. It is important here that
$\Gamma_\sigma$ does not intersect the weight lattice $\Lambda_W$; 
all this is explained in detail in the appendix. 

It is now easy to calculate the overlaps between two branes of the
form (\ref{ansatz}). We find
\begin{eqnarray}
\langle\!\langle  \sigma, \nu_1 |\!| \, q^{\frac{1}{2}(L_0+\bar{L}_0) -
\frac{c}{12}} \, |\!| \sigma, \nu_2 \rangle\!\rangle & = & 
|\Gamma_\sigma| \, 
\sum_{\lambda\in P^k_+} \, 
\frac{S_{\lambda \nu_1}^\ast S_{\lambda \nu_2}}{S_{\lambda 0}} \, 
\sum_{\mu-\lambda\in\Lambda_R}{}^{\!\!\!\!\prime} \;
\chi_{[\lambda,\mu]}(\tau) \, \chi^\h_{\mu}
(\tau) \nn \\
& = & \sum_{\lambda\in P^k_+} \, 
\frac{S_{\lambda \nu_1}^\ast S_{\lambda \nu_2}}{S_{\lambda 0}} \, 
\sum_{v\in\Gamma_\sigma}  \chi_\lambda (\tau,v,0)\,,
\end{eqnarray}
where $\chi_\lambda (\tau,v,t)$ is the unspecialised affine character,
\begin{equation}
\chi_\lambda (\tau,v,t) = e^{-2\pi {\rm i}\, t k}\, 
\hbox{Tr}_{\H_\lambda} 
\left( e^{2\pi {\rm i}\, \tau (L_0 - \frac{c}{24})} \, 
e^{2 \pi {\rm i}\, v\cdot H_0} \right) \,.
\end{equation}
Under a modular transformation, the unspecialised characters
$\chi_\lambda(\tau,v,0)$ 
transform into the characters $\chi_{\lambda'+v}$
of representations $\lambda'+v$, which are {\it twisted} by inner twists
associated to $v$ \cite{kacpeterson,kac}:
\begin{eqnarray}\label{modular}
\chi_\lambda(-1/\tau,v,0)  & =  &
\sum_{\lambda'\in P^k_+} \, S_{\lambda \lambda'} \,
\chi_{\lambda'} (\tau, \tau \, v, - \tau k v^2 /2) \nn \\
& = & \sum_{\lambda'\in P^k_+} \, S_{\lambda \lambda'} \,
\chi_{\lambda'+v} (\tau,0,0) \,.
\end{eqnarray}
(For a clear introduction to twisted
representations see for example \cite{go}.) The above cylinder diagram
therefore becomes in the open string channel
\begin{eqnarray}\label{openspectrum}
\langle\!\langle  \sigma, \nu_1 |\!| \, q^{\frac{1}{2}(L_0+\bar{L}_0) -
\frac{c}{12}} \, |\!| \sigma, \nu_2 \rangle\!\rangle & = & 
\sum_{\lambda\in P^k_+} \, 
\frac{S_{\lambda \nu_1}^\ast S_{\lambda \nu_2}}{S_{\lambda 0}} \, 
\sum_{v\in\Gamma_\sigma} 
\sum_{\lambda'\in P^k_+} \, S_{\lambda \lambda'} \,
\chi_{\lambda'+v} (\tau,0,0)  \nn \\ 
& = & \sum_{\lambda'\in P^k_+} \, N_{\nu_2 \lambda'}{}^{\nu_1} \, 
\sum_{v\in\Gamma_\sigma} \chi_{\lambda'+v} (\tau,0,0) \,,
\end{eqnarray}
where $N_{\nu_2 \lambda'}{}^{\nu_1}$ are the fusion rules of 
$\g_k$. In particular, this implies that these boundary states satisfy
Cardy's condition.
Since $\Gamma_\sigma$ does not intersect the weight lattice, 
$\chi_{\mu+v} (\tau,0,0)=\chi_{\mu'+v'}(\tau,0,0)$ if and only if 
$\mu=\mu'$ and $v=v'$. Thus the different representations that appear
on the right hand side of (\ref{openspectrum}) are in fact all
different.  

These boundary states therefore still define a NIM-rep
(for an introduction into these matters see for example
\cite{ggzero}), the only difference being that now the non-negative
matrices are also associated to twisted representations of
$\g_k$. Indeed, the above formula simply implies that for any
$\lambda$ that differs by a twist in $\Gamma_\sigma$ from 
$\lambda'\in P^k_+$  
\begin{equation}\label{nimtwist}
{\cal N}_{\nu_2 \lambda}{}^{\nu_1} = 
N_{\nu_2 [\lambda]}{}^{\nu_1} \,,
\end{equation}
where $[\lambda]$ is the unique untwisted representation that can be
obtained from $\lambda$ by a twist $v\in\Gamma_\sigma$. The fusion of
two twisted representations $\lambda$ and $\nu$ both of whose twists
are inner, is simply the fusion product of $[\lambda]$ and $[\mu]$, 
twisted by the sum of the two twists \cite{gabtwis}. Thus it is
manifest that  (\ref{nimtwist}) still defines a NIM-rep. (In fact,
this NIM-rep is simply the tensor product of the original NIM-rep
of $\g$, and the NIM-rep associated to $\Gamma_\sigma$.) Since all
twists in question are inner, the dimension that should be associated
to a twisted representation $\lambda$ is simply the dimension of the
corresponding untwisted representation $[\lambda]$. One can thus use
the same  arguments as in \cite{fs} to conclude that the charges that
are associated to these D-branes must satisfy 
\begin{equation}
\dim([\lambda]) \, q_{\nu} = \sum_{\nu'}
{\cal N}_{\nu \lambda}{}^{\nu'} \, q_{\nu'} \,.
\end{equation}
Since the above NIM-rep agrees with the fusion rule, the same analysis
as in \cite{fs} applies, and we conclude that each such family of
D-branes contributes one summand $\Zop_{d(G,k)}$ to the charge group.

\subsection{The twisted construction}

The analysis in the twisted case, \ie\ when we choose $\omega=C$ in
(\ref{cosetglue}), is very similar. In this case we get an Ishibashi
state for every $(\lambda,\mu,\bar\mu)$ provided that 
\begin{equation}
\lambda = \lambda^\ast \,, \qquad \qquad 
\mu = - \sigma\, \bar\mu \,, \qquad \qquad
\mu = + \bar\mu \qquad (\hbox{mod}\, k\Lambda_R) \,.
\end{equation}
If we define $\hat{\sigma}=-\sigma$, the last two conditions become 
\begin{equation}\label{condunt}
({\bf 1} - \hat{\sigma} ) \, \bar\mu \in k \Lambda_R \,,
\end{equation}
and therefore agree formally with (\ref{condun}). However,
$\hat{\sigma}$ differs by a sign from $\sigma$, and thus also the last
sign of $\hat{\sigma}$, $\hat{s}_n$, is opposite to the last sign of
$\sigma$, $s_n$. In particular, the choice $s_n=-1$ now leads to
$\hat{s}_n=+1$; in this case, the analysis of the solution space to
(\ref{condunt}) is then identical to that of the previous section. If
we tried to perform this construction for $s_n=+1$ we would run into
the same difficulties as in the previous case with $s_n=-1$. In
particular, as is explained in the appendix, it is then in general not
possible to choose $\Gamma_\sigma$ such that it does not intersect the
weight lattice. In the following we therefore assume that $s_n=-1$ so
that $\hat{s}_n=+1$.  

In addition to the condition on $\bar\mu$ we now only get Ishibashi
states that come from a self-conjugate $\lambda$. Thus the natural
ansatz for our boundary states is now 
\begin{equation}\label{ansatzt}
|\!| \sigma,x \rangle \!\rangle = \sqrt{|\Gamma_{\hat{\sigma}}|}\,
\sum_{\lambda\in \E_\omega} 
\frac{\hat{S}_{\lambda x}}{\sqrt{S_{\lambda 0}}} \,
\sum_{\bar\mu+\lambda\in\Lambda_R}{}^{\!\!\!\!\prime}\, 
e^{{\rm i}\, \theta \cdot   \bar\mu} 
\, |\lambda,\bar\mu\rangle\!\rangle \,,
\end{equation}
where the last sum is restricted to the solutions of (\ref{condunt})
and $|\Gamma_{\hat{\sigma}}|$ is the order of the
finite abelian group  
$\Gamma_{\hat{\sigma}}$ that was defined before. Here $\E_\omega$ is
the set of charge-conjugation invariant weights in $P^k_+$, and  
$\hat{S}_{\lambda x}$ is the twisted $S$-matrix that appears in the
construction of the twisted D-branes (see for example
\cite{ggzero}). These D-branes are now labelled by the twisted
representations $x$ of $\g_k$. Using the same calculation as in the
previous section we then find that these boundary states define indeed
a NIM-rep, and that it is simply given by 
\begin{equation}
{\cal N}_{x_2 \lambda}{}^{x_1} = 
{\cal N}^C_{x_2 [\lambda]}{}^{x_1} \,,
\end{equation}
where $[\lambda]$ is the unique untwisted representation that can be
obtained from $\lambda$ by a twist $v\in\Gamma_{\hat{\sigma}}$, and 
the matrix ${\cal N}^C_{x_2 \lambda}{}^{x_1}$ is the usual twisted  
NIM-rep of $\g_k$ (see for example \cite{ggzero}). Using the results
of \cite{gg} it then follows again that  each such family of D-branes
contributes one summand $\Zop_{d(G,k)}$ to the charge group.

\section{Concluding remarks}
\setcounter{equation}{0}

In the previous section we have constructed one set of D-branes for
each orthogonal real matrix $\sigma$ that is diagonal in the
basis defined by $\tilde\Lambda_j$. Depending on the value 
$s_n=\pm 1$, the construction resulted in the untwisted or twisted 
NIM-rep. Since the twisted NIM-rep gives rise to the same charges as
the untwisted NIM-rep \cite{gg}, each of these constructions leads to
a summand $\Zop_{d(G,k)}$ in the charge group. There are $\rk(\bar\g)$
different signs in the definition for $\sigma$, and thus we get in
total $2^{\rk(\bar\g)}$ constructions. 

Not all of these choices are inequivalent though. There is precisely
one non-trivial element of the Weyl group of $\bar\g$ for
which all $\tilde\Lambda_i$ are eigenvectors: it is the
Weyl-reflection corresponding to $\alpha_1$ which acts as $-1$
on  $\tilde\Lambda_1$, while leaving all other $\tilde\Lambda_i$ 
invariant. This Weyl transformation therefore maps a brane to one
with opposite $s_1$; strictly speaking, it also modifies the gluing
conditions on the coset algebra, but it seems very plausible that two
branes that have the same gluing condition on $\h$ will in fact carry
the same charge. Thus there are in fact only 
$2^{\rk(\bar\g)-1}$ branes with different charges, each of which leads
to a summand of $\Zop_{d(G,k)}$. This accounts precisely for the
charges that were determined using K-theory arguments.

It seems very difficult to prove rigorously that two given branes carry
different charges, \ie\ lie on different sheets of the moduli
space. However, it seems plausible that branes that have different
gluing conditions on $\h$ (up to the Weyl symmetry mentioned before)
are in fact inequivalent. [After all, in order to preserve
supersymmetry the fermions have to satisfy the corresponding gluing
conditions, and these boundary states therefore couple to different RR
ground states.] It would nevertheless be very interesting to 
establish this from first principles (for example by finding a
numerical invariant that characterises the different sheets). 

It would also be interesting to understand the geometrical
interpretation of these branes. We have found another realisation
\cite{ggr} of these D-brane charges along the lines of \cite{quella},
for which the geometrical interpretation is more obvious. We believe
that these D-branes describe another point of each sheet of the moduli
space.

\section*{Acknowledgments}

We thank Stefan Fredenhagen and Thomas Quella for useful
conversations. This work was completed while MRG and TG were visiting  
BIRS; we are very grateful for the wonderful working 
environment we experienced there! The research of MRG is also
supported in part by the Swiss National Science Foundation, while that
of TG is supported in part by NSERC.

\appendix

\section{The projector}
\setcounter{equation}{0}

In this appendix we supply some technical details, completing the
argument of section~3.

Recall that in our construction we retain only those weights
$\bar{\mu}$ obeying (\ref{3.17}) for all $j=1,\ldots,n$. Define   
\begin{equation}
v_j=\frac{1}{k} \left( (1-s_j)\alpha_j
+(s_{j-1}-s_j)\tilde{\Lambda}_{j-1} \right)\,. 
\end{equation}
Then for each $j$ define the projector
\begin{equation}
P_j=\frac{1}{L_j}\sum_{\ell=1}^{L_j}\exp[2\pi {\rm i}\, \ell v_j\cdot 
\bar{H}_0]\,,
\end{equation}
where 
\begin{equation}
L_j=\left\{
\begin{array}{cl}
kj & \hbox{if $s_j\ne s_{j-1}$} \\
k  & \hbox{if $s_j=s_{j-1}=-1$} \\
1  & \hbox{if $s_j=s_{j-1}=+1$,}
\end{array}
\right.
\end{equation}
and $\bar{H}_0$ is the operator 
$\bar{H}_0|\mu,\bar{\mu}\rangle
=\bar{\mu}|\mu,\bar{\mu}\rangle$. Clearly,
$P_j|\mu,\bar{\mu}\rangle=|\mu,\bar{\mu}\rangle$ or $0$,
depending on whether or not $\bar{\mu}$ satisfies the $j$th
equation (\ref{3.16}). Thus the desired projector (\ref{condun2})  is  
$P_\sigma=\prod_j P_j$. 

We can describe $P_\sigma$ additively as follows. Consider the
quotient $\Gamma_\sigma =\Lambda_v/(\Lambda_v\cap \Lambda_R)$ of the  
$\Zop$-span $\Lambda_v$ of the vectors $v_j$, by its intersection with
the root lattice $\Lambda_R$. Then $\Gamma_\sigma$ is a finite abelian 
group, and $P_\sigma$ can be written as in (\ref{condun2}).

Establishing the NIM-rep property in section~3 required that we verify
that $P_\sigma$ produces only twisted representations. More precisely,
we need to verify that, for any $v\in\Lambda_v$, $v\in\Lambda_W$ (the 
weight lattice) implies $v\in\Lambda_R$. This is easy to see. Recall
that it suffices to  consider $s_n=+1$. Suppose
$v\in\Lambda_v\cap\Lambda_W$. Then $v\in\Lambda_R$ if and only if
$\Lambda_n\cdot v\in\Zop$. We compute 
\begin{equation}
\sum_j\ell_jv_j\cdot\Lambda_n=\frac{\ell_n\,(1-s_n)}{k}=0\,,
\end{equation}
and the claim follows.

The order of the actual group $\Gamma_\sigma$ is a factor of the
product of the $L_j$. In order to compute the actual order of the
group $\Gamma_\sigma$, we observe that a linear combination 
$\sum_i \ell_i v_i$ is trivial in $\Gamma_\sigma$ if and only if it is
an element of $k\Lambda_R$. In particular, it must therefore satisfy 
\begin{equation}
\Lambda_j \cdot \left(\sum_i \ell_i v_i\right) \in k \Zop 
\end{equation}
for each $j$. It is easy to see that this leads to a triangular set of
conditions in the $\ell_i$. In particular, for large $k$ the order of
$\Gamma_\sigma$ grows like
\begin{equation}
|\Gamma_\sigma| \sim k^{n-m}\,,
\end{equation}
where $m$ is the number of $j$ for which $s_{j-1}=s_j=+1$, 
$j=1,\ldots, n$ with $s_0=+1$.

Incidentally, this is the place where we see explicitly that the   
construction of section~3 collapses when $s_n=-1$ (or equivalently if
we do not choose the gluing condition (\ref{cosetglue}) for the coset
algebra depending on the value of $s_n$). When $s_n=-1$, the
requirement $v\in\Lambda_R$ demands that $2\ell_n\in k\Zop$. This
will not be satisfied in general. For a concrete example take $n$ even
and $\sigma={\rm diag}(+1,+1,\ldots,+1,-1)$; $v=\ell_nv_n$ will lie in 
$\Lambda_v\cap \Lambda_W$ when $k$ divides $2\ell_n (n+1)/n$, which
certainly does not force $k$ to divide $2\ell_n$. Thus, when $s_n=-1$,
the construction will typically fail to produce a NIM-rep.

\end{document}